\documentclass[referee]{aa}
\usepackage{epsfig}
\usepackage{amssymb}
\def\etal{{\it et al.}}
\def\Journal#1#2#3#4{{#1} {\bf #2}, #3 (#4)}

\def\APJ{\em ApJ.}
\def\AST{\em Astron. J.}

\def\JPG{\em J. Phys. G: Nucl. Part. Phys.}

\def\PRL{\em Phys. Rev. Lett.}
\def\PRD{{\em Phys. Rev.} D}

\def\be{\begin{equation}}
\def\ee{\end{equation}}
\def\bea{\begin{eqnarray}}
\def\eea{\end{eqnarray}}
\begin{document}
   \thesaurus{12     % A&A Section 6: Form. struct. and evolut. of stars
             (12.04.1;  % Cosmogony,
              12.03.04;
              02.05.1)} % Stars: structure of.

\title{Cosmic Equation of State, Quintessence and Decaying Dark Matter}
\author{Houri Ziaeepour}
\institute{ESO, Schwarzchildstrasse 2, 85748, Garching b. M\"{u}nchen, Germany
\footnote{Present Address: 03, impasse de la Grande Boucherie,
F-67000, Strasbourg, France.}}

\date{Received ......; accepted ......}
\maketitle

%\begin{center}
%\Large \bf {Cosmic Equation of State, Quintessence and Decaying Dark Matter\\}
%\end{center} 

%\begin{center}
%{\it Houri Ziaeepour\\
%{ESO, Schwarzchildstrasse 2, 85748, Garching b. M\"{u}nchen, Germany
%\footnote{Present Address: 03, impasse de la Grande Boucherie,
%F-67000, Strasbourg, France.}\\
%Email: {\tt houri@eso.org}}}
%\end{center}

\begin {abstract}
If CDM particles decay and their lifetime is comparable to the age of the 
Universe, they can modify its equation of state. By comparing the results 
of numerical simulations with high redshift SN-Ia observations, we show that 
this hypothesis is consistent with present data. Fitting the simplest 
quintessence models with constant $w_q$ to the data leads to 
$w_q \lesssim -1$. We show that a universe 
with a cosmological constant or quintessence matter with $w_q \sim -1$ and 
a decaying Dark Matter has an effective $w_q < -1$ and fits SN data better 
than stable CDM or quintessence models with $w_q > -1$.
\end {abstract}

\begin{figure}[t]
\begin{center}
%\rule{5cm}{0.2mm}\hfill\rule{5cm}{0.2mm}
%\vskip 2.5cm
%\rule{5cm}{0.2mm}\hfill\rule{5cm}{0.2mm}
\psfig{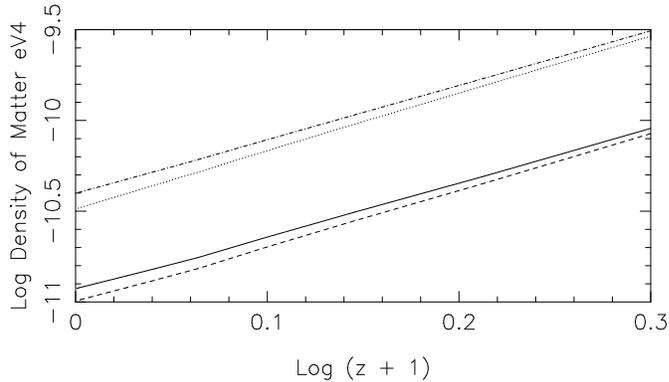}
\caption{Energy density of the Universe. Solid line $\Omega_{\Lambda}^{eq} = \Omega_{\Lambda} = 0.7$ and stable DM; dashed line the same cosmology with $\tau = 5 \tau_0$; dash dot line $\Lambda = 0$ and stable DM; dot line $\Lambda = 0$ and $\tau = 5 \tau_0$. Dependence on the mass of DM is negligible.\label{fig:t00}}
\end{center}
\end{figure}

\begin{figure}[t]
\begin{center}
%\rule{5cm}{0.2mm}\hfill\rule{5cm}{0.2mm}
%\vskip 2.5cm
%\rule{5cm}{0.2mm}\hfill\rule{5cm}{0.2mm}
\psfig{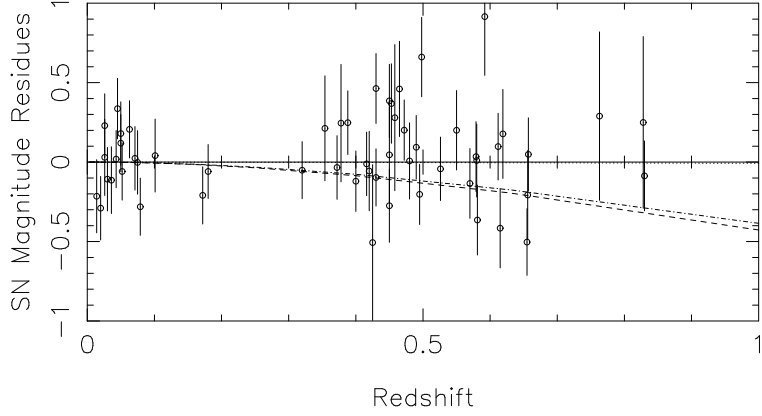}
\caption{Best fit residues with $\Omega_{\Lambda}^{eq} = 0.7$, $\tau = 5 
\tau_0$. It leads to $\Omega_{\Lambda} = 0.73$. The curves correspond to 
residue for stable DM with $\Omega_{\Lambda}^{eq} = \Omega_{\Lambda} = 0.7$ 
(doted); $\Lambda = 0$ and $\tau = 5 \tau_0$ (dashed); $\Lambda = 0$, stable 
DM (dash-dot).\label{fig:bestres}}
\end{center}
\end{figure}

\begin{figure}[t]
\begin{center}
%\rule{5cm}{0.2mm}\hfill\rule{5cm}{0.2mm}
%\vskip 2.5cm
%\rule{5cm}{0.2mm}\hfill\rule{5cm}{0.2mm}
\psfig{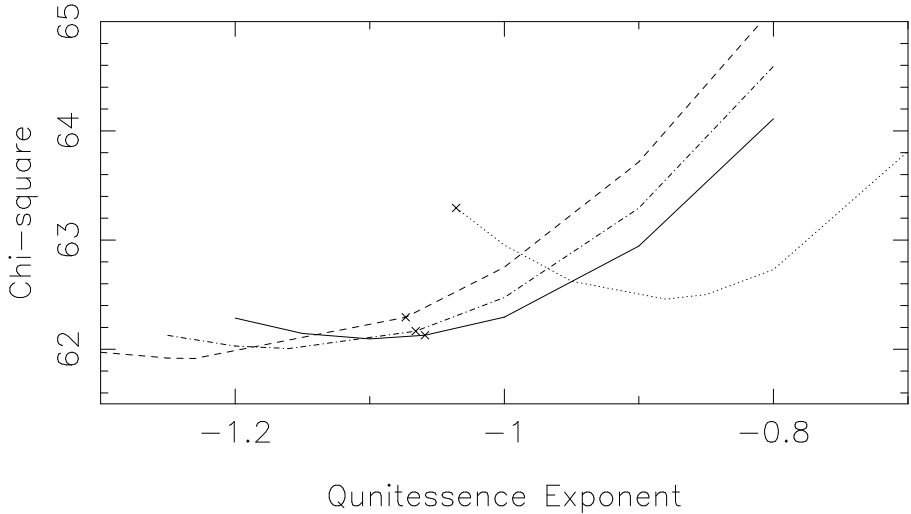}
\caption{$\chi^2$-fit of models defined in (\ref{quinanal}) as a function of 
$w_q$ for $\Omega_q = 0.67$ (dashed), $\Omega_q = 0.69$ (dash-dot), 
$\Omega_q = 0.71$ (solid) and $\Omega_q = 0.8$ (dotted). The $\chi^2$ of 
equivalent quintessence models to DDMs with $\tau = 5 \tau_0$ and same $\Omega_q$ is also shown. Except $\Omega_q = 0.8$ model, others are all the best fit 
to DDM. For $\Omega_q = 0.8$, a stable DM fits the data better, but the fit is 
poorer than former models.\label{fig:quindata}}
\end{center}
\end{figure}

There are at least two motivations for the existence of a Decaying Dark Matter 
(DDM). If R-parity in SUSY models is not strictly conserved, the LSP which 
is one of the best candidates of DM can decay to Standard Model 
particles Banks \etal~\cite{banks}. Violation of this symmetry is one of the 
many ways 
for providing neutrinos with very small mass and large mixing angle.\\
Another motivation is the search for sources of Ultra High Energy Cosmic 
Rays (UHECRs)(see Yoshida \& Dai~\cite {crrev} for review of their 
detection and Blandford~\cite{stdsrc} and Bhattacharjee \& Sigl~\cite{revorg} 
respectively for conventional and exotic sources). In this case, DDM 
must be composed of ultra heavy particles with $M_{DM} \sim 10^{22}-10^{25} 
eV$. In a recent work Ziaeepour~\cite{wimpzilla} we have shown that the 
lifetime of 
UHDM (Ultra Heavy Dark Matter) can be relatively short, i.e. $\tau \sim 10 - 
100 \tau_0$ where 
$\tau_0$ is the age of the Universe. Here we compare the prediction of this 
simulation for the Cosmic Equation of State (CES) with the observation of high 
redshift SN-Ia.\\
For details of the simulation we refer the reader to 
Ziaeepour~\cite{wimpzilla}. In 
summary, the decay of UHDM is assumed to be like the hadronization of two 
gluon jets. The decay remnants interact with cosmic backgrounds, notably CMB, 
IR, and relic neutrinos, lose their energy and leave a high energy 
background of stable species e.i. $e^\pm$, $p^\pm$, $\nu$, $\bar \nu$, and 
$\gamma$. We solve the Einstein-Boltzmann equations to determine the energy 
spectrum of remnants. Results of Ziaeepour~\cite{wimpzilla} show that 
in a homogeneous universe, even the short lifetime mentioned above can not 
explain the observed flux of UHECRs. The clumping of DM in the Galactic Halo 
however limits the possible age/contribution. These parameters are degenerate 
and we can not separate them. For simplicity, we assume that CDM is entirely 
composed of DDM and limit the lifetime. Fig. \ref{fig:t00} shows the 
evolution of energy density $T^{00}(z) \equiv \rho (z)$ at low and 
medium redshifts in a flat universe with and without a cosmological constant. 
As expected, the effect of DDM is more significant in a matter dominated 
universe i.e. when $\Lambda = 0$. For a given cosmology, the lifetime of 
DDM is the only parameter that significantly affects the evolution of $\rho$. 
For the same lifetime, the difference between $M_{DM} = 10^{12} eV$ and 
$M_{DM} = 10^{24} eV$ cases is only $\approx 0.4\%$. Consequently, in the 
following we neglect the effect of the DM mass.
For the same cosmological model and initial conditions, if DM decays, matter 
density at $z = 0$ is smaller than when it is stable because decay remnants 
remain highly relativistic even after losing part of their energy. Their 
density dilutes more rapidly with the expansion of the Universe than CDM and 
decreases the total matter density. Consequently, relative 
contribution of cosmological constant increases. This process mimics a 
quintessence model i.e. a changing cosmological constant Peebles \& 
Ratra~\cite{quinorg}, Zlatev, Wang \& Steinhardt~\cite{tracker} (see Sahni 
\& Starobinsky~\cite{quinrev} for recent review). However, the equation of 
state of this model has an exponent $w_q < -1$ which is in contrast with the 
prediction of scalar field models with positive potential (see appendix for 
an approximative analytical proof).\\
The most direct way for determination of cosmological densities and equation 
of state is the observation of SN-Ia's as standard candles. It is 
based on the measurement of apparent magnitude of the maximum of SNs 
lightcurve Perlmutter \etal~\cite{snmeasur}~\cite{snproj}, Riess A. \etal
~\cite{snmeasur1}. After correction for various 
observational and intrinsic variations like K-correction, width-luminosity 
relation, reddening and Galactic extinction, it is assumed that their 
magnitude 
is universal. Therefore the difference in apparent magnitude is only related 
to difference in distance and consequently to cosmological parameters.\\
The apparent magnitude of an object $m (z)$ is related to its absolute 
magnitude $M$:
\be
m (z) = M + 25 + 5 \log D_L
\ee
where $D_L$ is the Hubble-constant-free luminosity distance:
\be
D_L = \frac {(z + 1)}{\sqrt{|\Omega_R|}} {\mathcal S} \biggl (\sqrt{|\Omega_R|} 
\int_{0}^{z} \frac {dz'}{E (z')}\biggr ) \label {Dl}\\
\ee
\be
{\mathcal S}(x) = 
%\begin {cases}
\begin {tabular}{ll}
$\sinh (x)$ & $\Omega_R > 0$,\\ 
$x$ & $\Omega_R = 0$,\\
$\sin (x)$ & $\Omega_R < 0$.
\end {tabular}
%\end {cases}
\ee
\bea
E (z) & = & \frac {H (z)}{H_0}. \label {ez}\\
H^2 (z) & = & \frac {8\pi G}{3} T^{00} (z) + \frac {\Lambda}{3}. \label {hz}
\eea
Here we only consider flat cosmologies.\\
We use the published results of the Supernova Cosmology Project, Perlmutter 
\etal~\cite{snproj} for high redshift and Calan-Tololo sample, Hamuy 
\etal~\cite{tololo} for low redshift 
supernovas and compare them with our simulation. From these data 
sets we eliminate 4 SNs with largest residue and stretch as explained in 
Perlmutter \etal~\cite{snproj} (i.e. we use objects used in their fit B).\\
Minimum-$\chi^2$ fit method is applied to the data to extract the parameters 
of the cosmological models. In all fits described in this letter we consider 
$M$ as a free parameter and minimize the $\chi^2$ with respect to it. Its 
variation in our fits stays in the acceptable range of $\pm 0.17$, 
Perlmutter \etal~\cite{snmeasur}.\\
We have restricted our calculation to a range of 
parameters close to the best fit of Perlmutter \etal~\cite{snproj} i.e. 
$2.38 \times 10^{-11} \leqslant \rho_\Lambda \equiv \frac {\Lambda}{8\pi G} 
\leqslant 3.17 \times 10^{-11}eV^4$. The reason why we use $\rho_\Lambda$ 
rather than $\Omega_\Lambda$ is that the latter quantity depends on the 
equation of state and the lifetime of the Dark Matter. The range of 
$\rho_\Lambda$ given here is equivalent to $0.6\leqslant \Omega_{\Lambda}^{eq} 
\leqslant 0.8$ for a stable CDM and $H_0 = 70$ $km$ $Mpc^{-1} \sec^{-1}$ (we use 
$\Omega_{\Lambda}^{eq}$ notation to distinguish between this quantity and real 
$\Omega_\Lambda$).\\
Fig.\ref{fig:bestres} shows the residues of the best fit to DDM simulation. 
Although up to 
$1$-$\sigma$ uncertainty all models with stable or decaying DM with $5 \tau_0 
\lesssim \tau \lesssim 50 \tau_0$ and $0.68\lesssim \Omega_{\Lambda}^{eq} 
\lesssim 0.72$ are compatible with the data, a decaying DM with $\tau \sim 5 
\tau_0$ systematically fits the data better than stable DM with the same 
$\Omega_{\Lambda}^{eq}$. Models with $\Lambda = 0$ are ruled 
out with more than $99\%$ confidence level.\\
In fitting the results of DM decay simulation to the data we have directly 
used the 
equation (\ref{hz}) without defining any analytical form for the evolution of 
$T^{00}(z)$. It is not usually the way data is fitted to cosmological 
models Perlmutter \etal~\cite{snmeasur},~\cite{snmeasur1}, Garnavich 
\etal~\cite{coseq}. Consequently, we have also fitted an analytical 
model to the simulation for $z < 1$ as it is the redshift range of the 
available data. It includes a stable DM and a quintessence matter. Its 
evolution equation is:
\be
H^2 (z) = \frac {8\pi G}{3} (T^{00}_{st} + \Omega_q (z + 1)^{3 (w_q + 1)}). \label {quineq}
\ee
The term $T^{00}_{st}$ is obtained from our simulation when DM is stable. 
In addition to CDM, it includes a small contribution from hot components i.e 
CMB and relic neutrinos. For a given $\Omega_{\Lambda}^{eq}$ and 
$\tau$, the quintessence term is fitted to $T^{00} - T^{00}_{st} + 
\frac {\Lambda}{8\pi G}$.
\footnote{The exact equivalent model should have 
the same form as (\ref {quinanal}). However, it is easy to verify that in 
this case, the minimization of $\chi^2$ has a trivial solution with $w_q = 
-1$, $\Omega_q = 0$. Only one equation remains for non-trivial solutions and 
it depends on both $w_q$ and $\Omega_q$. Consequently, there are infinite 
number of solutions.\\
The model we have used here generates a very good equivalent model to DDM with 
less than $2\%$ error (Because CDM and quintessence terms are not fitted 
together, $\Omega$ is not exactly $1$).\label {foot}}
The results of this fit are $\Omega_q$ and $w_q$ 
which characterize an equivalent quintessence model for the corresponding 
DDM. The analytical model fits the simulation extremely good and the 
absolute value of relative residues is less than $0.2\%$. Results for models 
in the $1$-$\sigma$ distance of the best fit is summarized in 
Table \ref{tab:quineq}.\\ 
\begin{table}[t]
\caption{Cosmological parameters from simulation of a decaying DM and 
parameters of the equivalent quintessence model. $H_0$ is in $km$ $Mpc^{-1}
\sec^{-1}$.\label{tab:quineq}}
\vspace{0.2cm}
\begin{center}
\footnotesize
\begin{tabular}{|c|c|c|c|c|c|c|c|c|c|}
\hline
 &
\multicolumn {3}{c|}{Stable DM} & 
\multicolumn {3}{c|}{$\tau = 50 \tau_0$} & 
\multicolumn {3}{c|}{$\tau = 5 \tau_0$} \\
\hline
 &
\raisebox{0pt}[13pt][7pt]{$\Omega_{\Lambda}^{eq} = 0.68$} &
\raisebox{0pt}[13pt][7pt]{$\Omega_{\Lambda}^{eq} = 0.7$} &
\raisebox{0pt}[13pt][7pt]{$\Omega_{\Lambda}^{eq} = 0.72$} &
\raisebox{0pt}[13pt][7pt]{$\Omega_{\Lambda}^{eq} = 0.68$} &
\raisebox{0pt}[13pt][7pt]{$\Omega_{\Lambda}^{eq} = 0.7$} &
\raisebox{0pt}[13pt][7pt]{$\Omega_{\Lambda}^{eq} = 0.72$} &
\raisebox{0pt}[13pt][7pt]{$\Omega_{\Lambda}^{eq} = 0.68$} &
\raisebox{0pt}[13pt][7pt]{$\Omega_{\Lambda}^{eq} = 0.7$} &
\raisebox{0pt}[13pt][7pt]{$\Omega_{\Lambda}^{eq} = 0.72$}\\
\hline
\raisebox{0pt}[12pt][6pt]{$H_0$}  
 & \raisebox{0pt}[12pt][6pt]{$69.953$}
 & \raisebox{0pt}[12pt][6pt]{$69.951$} & \raisebox{0pt}[12pt][6pt]{$69.949$}
 & \raisebox{0pt}[12pt][6pt]{$69.779$} & \raisebox{0pt}[12pt][6pt]{$69.789$}
 & \raisebox{0pt}[12pt][6pt]{$69.801$} & \raisebox{0pt}[12pt][6pt]{$68.301$}
 & \raisebox{0pt}[12pt][6pt]{$68.415$} & \raisebox{0pt}[12pt][6pt]{$68.550$}\\
\hline
\raisebox{0pt}[12pt][6pt]{$\Omega_{\Lambda}$}
 & \raisebox{0pt}[12pt][6pt]{$0.681$}
 & \raisebox{0pt}[12pt][6pt]{$0.701$} & \raisebox{0pt}[12pt][6pt]{$0.721$}
 & \raisebox{0pt}[12pt][6pt]{$0.684$} & \raisebox{0pt}[12pt][6pt]{$0.704$}
 & \raisebox{0pt}[12pt][6pt]{$0.724$} & \raisebox{0pt}[12pt][6pt]{$0.714$}
 & \raisebox{0pt}[12pt][6pt]{$0.733$} & \raisebox{0pt}[12pt][6pt]{$0.751$}\\
\hline
\raisebox{0pt}[12pt][6pt]{$\Omega_q$}
 & \raisebox{0pt}[12pt][6pt]{-}
 & \raisebox{0pt}[12pt][6pt]{-} & \raisebox{0pt}[12pt][6pt]{-}
 & \raisebox{0pt}[12pt][6pt]{$0.679$} & \raisebox{0pt}[12pt][6pt]{$0.700$}
 & \raisebox{0pt}[12pt][6pt]{$0.720$} & \raisebox{0pt}[12pt][6pt]{$0.667$}
 & \raisebox{0pt}[12pt][6pt]{$0.689$} & \raisebox{0pt}[12pt][6pt]{$0.711$}\\
\hline
\raisebox{0pt}[12pt][6pt]{$w_q$}
 & \raisebox{0pt}[12pt][6pt]{-}
 & \raisebox{0pt}[12pt][6pt]{-} & \raisebox{0pt}[12pt][6pt]{-}
 & \raisebox{0pt}[12pt][6pt]{$-1.0066$} & \raisebox{0pt}[12pt][6pt]{$-1.0060$}
 & \raisebox{0pt}[12pt][6pt]{$-1.0055$} & \raisebox{0pt}[12pt][6pt]{$-1.0732$}
 & \raisebox{0pt}[12pt][6pt]{$-1.0658$} & \raisebox{0pt}[12pt][6pt]{$-1.0590$}\\
\hline
\raisebox{0pt}[12pt][6pt]{$\chi^2$}
 & \raisebox{0pt}[12pt][6pt]{$62.36$}
 & \raisebox{0pt}[12pt][6pt]{$62.23$} & \raisebox{0pt}[12pt][6pt]{$62.21$}
 & \raisebox{0pt}[12pt][6pt]{$62.34$} & \raisebox{0pt}[12pt][6pt]{$62.22$}
 & \raisebox{0pt}[12pt][6pt]{$62.21$} & \raisebox{0pt}[12pt][6pt]{$62.22$}
 & \raisebox{0pt}[12pt][6pt]{$62.15$} & \raisebox{0pt}[12pt][6pt]{$62.20$}\\
\hline
\end{tabular}
\end{center}
\end{table}
In the next step, we fit an analytical model to the SN-Ia data. Its evolution 
equation is the following:
\be
H^2 (z) = \frac {8\pi G}{3} ((1 - \Omega_q) (z + 1)^3 + 
\Omega_q (z + 1)^{3 (w_q + 1)}). \label {quinanal}
\ee
The aim for this exercise is to compare DDM equivalent quintessence models 
with the data.\\
Fig.\ref{fig:quindata} shows the $\chi^2$ of these fits as a function of 
$w_q$ for various values of $\Omega_q$. The reason behind using $\chi^2$ 
rather than confidence level is that it directly shows the goodness-of-fit. 
As with available data all relevant models are compatible up to $1$-$\sigma$, 
the error analysis is less important than goodness-of-fit and its behavior in 
the parameter-space.\\
Models presented in Fig.\ref{fig:quindata} have the same $\Omega_q$ as 
equivalent quintessence models obtained 
from DDM and listed in Tab.\ref{tab:quineq}. These latter models are shown 
too. In spite of statistical closeness of all fits, the systematic tendency 
of the minimum of $\chi^2$ to $w_q < -1$ when $\Omega_q < 0.75$ is evident. 
The minimum of models with $\Omega_q > 0.75$ has $w_q > -1$, but the fit is 
worse than former cases. Between DDM models, one with $\Omega_q = 0.71$ is 
very close to the best fit of (\ref{quinanal}) models with the same 
$\Omega_q$. 
Regarding errors however, all these models, except $\Omega_q = 0.8$ are 
$1$-$\sigma$ compatible with the data.\\
One has to remark that $\Omega_q$ and $w_q$ are not completely independent 
(see Footnote \ref{foot}) and models with smaller $\Omega_q$ and smaller 
$w_q$ has even smaller $\chi^2$. In fact the best fit corresponds to 
$\Omega_q = 0.5$, $w_q = -2.6$ with $\chi^2 = 61.33$. The rejection of these 
models however is based on physical grounds. In fact, if the quintessence 
matter is a 
scalar field, to make such a model, not only its potential must be negative, 
but also its kinetic energy must be comparable to the absolute value of the 
potential and this is in contradiction with very slow variation of the field. 
In addition, these models are unstable against perturbations. It is however 
possible to make models with $w_q < -1$, but they need unconventional kinetic 
term Caldwell~\cite {negqw}.\\ 
These results are compatible with the analysis performed by Garnavich 
\etal~\cite {coseq}. However, based on null energy 
condition Wald~\cite {wald}, they only consider models with $w_q \geqslant -1$. This 
condition should be satisfied by non-interacting matter and by total 
energy-momentum tensor. As our example of a decaying matter shows, a 
component or an equivalent component of energy-momentum tensor can have 
$w_q < -1$ when interactions are present.\\
In conclusion, we have shown that a flat cosmological model including a 
decaying dark matter with $\tau \sim 5 \tau_0$ and a cosmological constant 
or a quintessence matter with $w_q \sim -1$ at $z < 1$ and $\Omega_q \sim 
0.7$ fits the SN-Ia data better than models with a stable DM or $w_q > -1$.\\
The effect of a decaying dark matter on the Cosmic Equation of State (CES) 
is a distinctive signature that can hardly be mimicked by other phenomena, 
e.g. conventional sources of Cosmic Rays. It is an independent mean for 
verifying the hypothesis of a decaying UHDM. In fact if a decaying DM 
affects CES significantly, it must be very heavy. Our simulation of a decaying 
DM with $M \sim 10^{12} eV$ and $\tau = 5 \tau_0$ leads to an 
over-production by a few orders of magnitude of $\gamma$-ray background at 
$E \sim 10^{9}-10^{11} eV$ with respect to EGRET observation Sreekumar 
\etal~\cite{egret}. 
Consequently, such a DM must have a lifetime much longer than $5 \tau_0$. 
However, in this case it can not leave a significant effect on CES.\\
The only other alternative for making a 
quintessence term in CES with $w_q$ sightly smaller than $-1$, is a scalar 
field with a negative potential. Nevertheless, as most of quintessence models 
originate from SUSY, the potential should be strictly positive. Even if a 
negative potential or unconventional models are not a prohibiting conditions, 
they rule out a large number of candidates.\\
Present SN-Ia data is too scarce to distinguish with high precision between 
various models. However, our results are encouraging and give the hope that 
SN-Ia observations will help to better understand the nature of the Dark 
Matter in addition to cosmology of the Universe.

{\bf Appendix:} Here we use an approximative solution of (\ref {hz}) to find 
an analytical expression for the equivalent quintessence model of a cosmology 
with DDM and a cosmological constant. With a good precision the total density 
of such models can be written as the following:
\be
\frac {\rho (z)}{{\rho}_c} \approx {\Omega}_M (1 + z)^3 \exp 
(\frac {{\tau}_0 - t}
{\tau}) + {\Omega}_{Hot} (1 + z)^4 + {\Omega}_M (1 + z)^4 \biggl (1 - 
\exp (\frac {{\tau}_0 - t}{\tau}) \biggr ) + {\Omega}_{\Lambda}. \label {totdens}
\ee
We assume a flat cosmology i.e. ${\Omega}_M + {\Omega}_{\Lambda} = 1$ (ignoring the hot part). 
${\rho}_c$ is the present critical density. If DM is stable and we neglect the 
contribution of HDM, the expansion factor $a (t)$ is:
\be
\frac {a (t)}{a ({\tau}_0)} = \biggl [ \frac {(B \exp (\alpha (t - 
{\tau}_0)) - 1)^2}{4AB \exp (\alpha (t - {\tau}_0))}\biggr ]^{\frac {1}{3}} 
\equiv \frac {1}{1 + z}. \label {at}
\ee
\bea
A & \equiv & \frac {{\Omega}_{\Lambda}}{1 - {\Omega}_{\Lambda}}, \\
B & \equiv & \frac {1 + \sqrt {{\Omega}_{\Lambda}}}{1 - 
\sqrt {{\Omega}_{\Lambda}}}, \\
\alpha & \equiv & 3 H_0 \sqrt {{\Omega}_{\Lambda}}.
\eea
Using (\ref {at}) as an approximation for $\frac {a (t)}{a ({\tau}_0)}$ when 
DM slowly decays, (\ref {totdens}) takes the following form:
\bea
\frac {\rho (z)}{{\rho}_c} & \approx & {\Omega}_M (1 + z)^3 C^{-\frac {1}
{\alpha \tau}} + {\Omega}_{Hot} (1 + z)^4 + {\Omega}_M (1 + z)^4 (1 - 
C^{-\frac {1}{\alpha \tau}}) + {\Omega}_{\Lambda}. \label {totdens1}\\
C & \equiv & \frac {1}{B} \biggl (1 + \frac {4A}{(1 + z)^3} - \sqrt {(1 + \frac {4A}{(1 + z)^3})^2 - 1} \biggr).
\eea
For a slowly decaying DDM, $\alpha \tau >> 1$ and (\ref {totdens1}) becomes:
\bea
\frac {\rho (z)}{{\rho}_c} & \approx & {\Omega}_M (1 + z)^3 + {\Omega}_{Hot} 
(1 + z)^4 + {\Omega}_q (1 + z)^{3 {\gamma}_q}, \label {totdens2} \\
{\Omega}_q (1 + z)^{3 {\gamma}_q} & \equiv & {\Omega}_{\Lambda} (1 + 
\frac {{\Omega}_M}{\alpha \tau {\Omega}_{\Lambda}} z (1 + z)^3 \ln C). 
\label {qeqdef}
\eea
Equation (\ref {qeqdef}) is the definition of equivalent quintessence matter. 
After its linearization:
\be
w_q \equiv {\gamma}_q - 1 \approx \frac {{\Omega}_M (1 + 4 A)(1 - \sqrt {2 A})}
{3 \alpha \tau {\Omega}_{\Lambda} B} - 1.
\ee
It is easy to see that in this approximation $w_q < -1$ if ${\Omega}_{\Lambda} > \frac {1}{3}$.

\end {document}